\documentclass[12pt]{iopart}

\usepackage{graphicx}
\usepackage{iopams} 
\usepackage{color}
 
\begin{document}

\title[Polysomally Protected Viruses]{Polysomally Protected Viruses}

\author{Michael Wilkinson$^{1,2}$, David  Yllanes$^1$ and Greg Huber$^1$}

\address{$^1$ Chan Zuckerberg Biohub, 499 Illinois Street, San Francisco, CA 94158, USA\\
$^2$ School of Mathematics and Statistics, The Open University,Walton Hall, Milton Keynes, MK7 6AA, UK}

\ead{michael.wilkinson@czbiohub.org}
\vspace{10pt}
\begin{indented}
\item[]January 2021
\end{indented}

\begin{abstract}
It is conceivable that an RNA  virus could use a polysome, that is, a string of ribosomes covering the RNA strand, 
to protect the genetic material from degradation inside a host cell. This paper discusses how such a 
virus might operate, and how its presence might be detected by ribosome profiling. There are two possible 
forms for such a \emph{polysomally protected virus}, depending upon whether just the forward strand or 
both the forward and complementary strands can be encased by ribosomes (these will be termed type 1 
and type 2, respectively). It is argued that in the type 2 case the viral RNA would evolve an 
\emph{ambigrammatic} property, whereby the viral genes are free of stop codons in a reverse 
reading frame (with forward and reverse codons aligned). 
Recent observations of ribosome profiles of ambigrammatic narnavirus sequences are consistent 
with our predictions for the type 2 case.
\end{abstract}

\section{Introduction}

A canonical model for the structure of a virus \cite{Cob+16} consists of genetic material 
encased in a capsid composed of a protein shell. A simpler model has also been observed, termed a 
narnavirus (this term is a contraction of \lq naked RNA virus'). The narnavirus examples that have 
been characterised appear to be single genes, which code for an RNA-dependent RNA-polymerase
(abbreviated as RdRp) \cite{Hillman2013}. It appears to be advantageous to the 
propagation of a virus if the genetic material can be encapsulated at 
some stage in its replication cycle, and it appears natural to ask whether 
some very simple RNA viruses could co-opt part of the machinery of the host cell in order to 
build a container. The most natural candidate is to make a covering out of ribosomes, which 
already contain an internal channel that can bind to RNA. 
If viral RNA can be completely covered with a chain of ribosomes, it could be well 
protected from defence mechanisms of host organism, because the exterior of the package presents 
molecules which are part of the host cells. This paper discusses how a class of very simple viruses 
could make a container for their genetic material out of ribosomes, resulting in a class of RNA viral 
systems which are, in some sense, intermediate between narnaviruses and conventional viruses. 
The covering structure, consisting of a chain of ribosomes attached to the viral RNA, would be 
analogous to a polysome \cite{Nol08,Sin96,Chr+99,Bra+09}, and for this reason we shall refer to these systems 
as \lq polysomally protected viruses', abbreviated hereafter as PolyProV. Such structures could be 
reservoirs of viral RNA which can be protected from degradation and hidden from defence 
mechanisms that might detect viral RNA. These protected viruses can be propagated 
\lq vertically' by cell division. The possibility that viral RNA could be shielded by 
a layer of ribosomes was discussed in a recent popular article \cite{Cep20}, and is also 
mentioned in a recent preprint \cite{Ret+20}. It is the purpose of this work to discuss the 
mechanism by which this can be realised, and the how it can be detected by ribosome profiling.

\begin{figure}
\begin{center}
\includegraphics[width=0.85\textwidth]{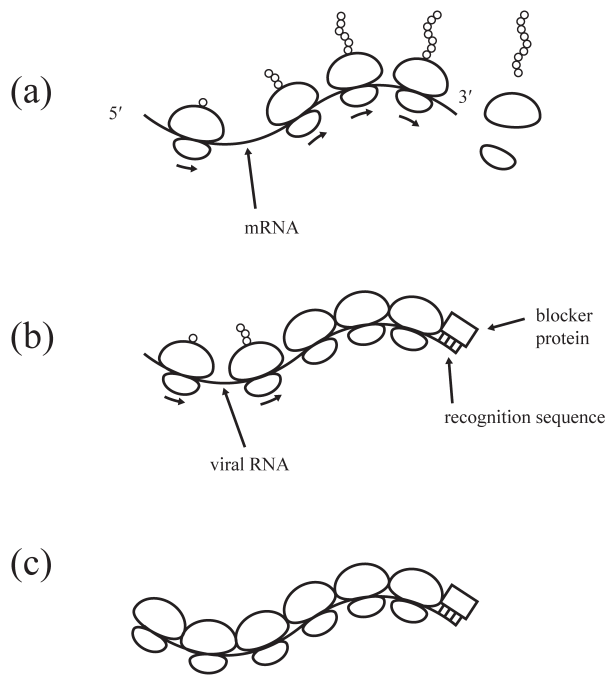}
\caption{\label{fig: 1}
{\bf (a)} A polysome consists of a number of ribosomes attached to a RNA molecule (usually mRNA). The ribosomes 
attach to the $5'$ end and move along the RNA, translating polypeptide chains as they go. {\bf (b)}
Our hypothetical polysomally protected virus is an RNA virus system including a gene that creates a \lq blocking' 
macromolecule (either a protein or possibly an RNA segment), 
which binds to a recognition site at the $3'$ end of the virus.  
Ribosomes are able to attach to the virus RNA at the 
$5'$ end, but are not released at the $3'$ end. {\bf (c)} The viral RNA becomes coated in ribosomes, which 
are frozen into fixed positions, and which form a protective sheath. 
}
\end{center}
\end{figure}

A conventional polysome is an open system where ribosomes move along the RNA 
chain~\cite{Nol08,Sin96,Chr+99,Bra+09},
synthesising a polypeptide chain as they go, and eventually detach from one end, see figure~\ref{fig: 1}({\bf a}), or
when they encounter a stop codon. 
The type of encapsulation that we propose is one where the ribosomes are stuck in position. 
This means that we must hypothesise a mechanism which creates the polysome shell by preventing 
ribosomes from detaching from the $3'$ end of the viral RNA (figure \ref{fig: 1}({\bf b})), thus 
creating a \lq frozen' polysome.  We propose that ribosomes attach to the $5'$ end of the viral RNA 
chain and move along the RNA chain until they form a string of ribosomes which are in close contact, 
like a string of pearls (figure \ref{fig: 1}({\bf c})).

Cells have machinery to release ribosomes which are not functioning efficiently \cite{Joa17}. In particular, 
\lq stalled' ribosomes are released by a process known as \lq no-go decay', abbreviated as NGD, 
which is an active field of study \cite{Mat+17,Ren+18,Jus+18,Nav+20}. The polysomally protected 
virus system would have to either disrupt the NGD process, or else infect cells where this process is defective. 
Given the complexity of the machinery required to implement \lq no-go decay', it must have many 
potential vulnerabilities.

We can imagine two forms of this class of virus. In its simplest form, termed PolyProV1, a polysomal sheath 
is only able to cover the forward strand of the RNA. Creation of a complementary strand is a necessary 
part of the replication cycle of the viral RNA, and in the simplest form, the complementary strand 
is not protected. We can also propose that there exists a type of this viral system, denoted by PolyProV2, 
where both the forward and complementary strands can be protected by being encased in a chain of 
ribosomes.

We discuss what would be the characteristic properties of such a system, and how their presence might 
be detected. Both types, PolyProV1 and PolyProV2, may show distinctive signatures 
under \lq ribosome profiling' (see \cite{Ingolia2009,Ingolia2012,Bra+15,Ingolia2016} for a discussion of this technique), 
and we give an indication of what might be expected. We remark that recent 
experiments on a narnavirus system \emph{Culex narnavirus 1}, reported in \cite{Ret+20}, show 
precisely the type of ribosome profile signatures that we describe, without explaining their form. (Figure 3
of their paper shows the phenomenon that we explain in section \ref{sec: 3} below, leading to distinctive 
profile features illustrated schematically in our figures \ref{fig: 3} and \ref{fig: 4}).
We also argue that, in the case of PolyProV2 systems, 
there would be a very distinctive signature in the genetic code of the virus. The formation of a polysome 
which covers the whole of the strand requires that the RNA sequence should not have any stop codons
(that is, it should have an open reading frame, abbreviated as ORF). The genes of the PolyProV2 system
would therefore have to have a reading frame which is devoid of stop codons on the complementary strand, 
as well as the forward strand. We have previously discussed the evolution of genetic sequences 
which are \lq ambigrammatic', that is, readable in both forward and reverse directions, showing that 
stop codons in the reverse-read direction can be eliminated even if the amino-acid sequence of a 
gene is strictly conserved \cite{DeR+19}. We argue that a recent observation of two  
ambigrammatic sequences in the \emph{Culex narnavirus 1} system reported in \cite{Bat+20,Ret+20} is a 
very strong candidate to be a PolyProV2 type virus.

Ambigrammatic sequences have been observed in narnavirus systems 
\cite{Cook2013,DeR+19,Cep20,Bat+20,Ret+20,Dudas2021}, and it is possible 
that the ORF on the complementary strand code for a functional protein. In a separate 
paper \cite{Dudas2021} we shall discuss criteria based upon statistical studies of polymorphism 
which could distinguish the PolyProV2 system from a narnavirus which has a functional gene 
on the complementary strand. Our results for  \emph{Culex narnavirus 1} and for \emph{Zheijiang mosquito virus 3}
indicate that the complementary strands do not code for a functional protein.
These two likely candidates for polysomally protected viruses are both narnaviruses, which along with 
viroids and virusoids \cite{Sym91}, are the simplest infectious agents. However, the device of using a covering of 
ribosomes to create a reservoir of viral RNA to facilitate vertical transmission is something which could be 
adopted by more sophisticated viruses. Polysomal protection may eventually be found to be a commonly 
occurring mode of virus propagation.

Most theoretical studies of polysomes have emphasised models based upon the totally 
asymmetric exclusion process~\cite{Lakatos2003,Lakatos2004,Lakatos2005,DaoDuc2018,Erdmann2020}, 
and some of these papers have considered phases where there are \lq traffic jams' formed by slowly 
moving ribosomes. Our theory considers a quite different phenomenon, where the ribosomes are \emph{stationary} 
because their release at the $3'$ end is blocked.

\section{Predicted properties}
\label{sec: 2}

Let us assume that an RNA virus does use a \lq frozen' polysome to create a 
covering out of ribosomes, and consider what are the plausible consequences of this hypothesis.
There are two questions that should be addressed. Firstly, how is the frozen polysome
created? And secondly, is it possible to protect the complementary strand as well as the 
coding strand of the virus?

\subsection{Creating the polysomal sheath}
\label{sec: 2.1}

The most natural hypothesis about the mechanism to create frozen polysomes 
is that ribosomes are prevented from detaching from the $3'$ end of the RNA.
The simplest mechanism for this is for there to exist a macromolecule (a protein, or an RNA segment) 
which binds to the $3'$ end of the viral RNA to block ribosomes from detaching. 
At least one gene would be required to code for this \lq end-stop' macromolecule.

The mechanism which freezes polysomes must have a quite 
specific switch, which can distinguish virus RNA from the host mRNA (if this were 
not the case, then the \lq end-stop' would inhibit all translation processes indiscriminately, 
damaging the host cell). The required specificity would have to be achieved by a signalling 
sequence in the virus RNA, such that the end-stop only binds when the signalling 
sequence is present. The only plausible location for the signal sequence is at the $3'$ end of the virus 
RNA chain, where the end-stop protein will bind.

These considerations imply that the simplest polysomal virus would have two genes, 
one to make the RdRp to replicate the virus, and the other one to make a blocking 
molecule to stop ribosomes from detaching from the $3'$ end of the viral RNA. In addition, there
must be a recognition sequence at the $3'$ end of the virus chain. Eukaryotic cells 
have mechanisms for releasing \lq stalled' ribosomes \cite{Joa17,Mat+17,Ren+18,Jus+18,Nav+20},
and if polysomally protected viruses exist, that may be associated with other virus genes 
which disrupt the mechanisms which release stalled ribosomes.

It is important to note that this picture implies a mechanism whereby the ribosomes are switched 
from replicating more viral RNA to acting as a shield. In the initial stages of infection of a cell, the RdRp 
will replicate viral RNA freely. The products of this process will include the molecule which binds to the 
viral RNA, having the effect of blocking further transcription. As the viral load inside a cell increases, the 
blocker molecules bind to the $3'$ ends of viral RNA creating a reservoir of viral RNA which is protected 
from degradation. 

This mechanism creates a reservoir of viral RNA inside a cell which is protected from degradation by being 
covered by ribosomes. There must, in turn, be a route for the protected viral RNA to become active again. The simplest 
possibility is that the binding of the blocker molecule to the $3'$ end is reversible, and so that transcription 
of viral RNA re-commences when the concentration of the blocker molecule decreases.

\subsection{The ambigrammatic advantage: protecting the complementary strand}
\label{sec: 2.2}

Replication of the virus RNA by the RdRp requires making a complementary copy. 
In addition to protecting the coding strand of the RNA, the polysomal virus could 
also evolve so that the complementary strand can be protected. Let us consider the additional 
features that are required to convert a PolyProV1 system, where just one strand is protected, 
to a PolyProV2 system, where both the forward and the complementary strands can be 
enclosed by a frozen polysome.

For a typical RNA sequence there will be stop codons on the complementary strand which 
would cause ribosomes to detach, preventing the RNA sequence from becoming shielded 
inside a polysome. This can be avoided if the RNA sequence is \emph{ambigrammatic}, 
in the sense that it is readable, without encountering stop codons, in both a forward and reverse 
reading frame. Recently, it has been shown that it is always possible to create an ambigrammatic sequence 
by substitution of codons by synonyms \cite{DeR+19}. This mechanism gives a rapid route to 
evolving an ambigrammatic sequence, without detriment to the function of 
genes translated in the forward direction. The ORF for the complementary strand 
must have the its codons aligned with the ORF for transcription on the forward strand \cite{DeR+19}.

There is an additional requirement for the complementary strand to be protected: 
the $3'$ end of the complementary strand has to have a recognition sequence 
to signal the end-stop protein to attach itself. This implies that there is a reverse 
complement of a valid recognition sequence at the $5'$ end of the coding strand. 
The simplest implementation of this is if the $5'$ end has a sequence which is the 
reverse complement of the recognition segment at the $3'$ end of the coding strand.

In this context, we remark that narnaviruses typically have a sequence CCCC at the $3'$ end, 
and a complementary sequence GGGG at the $5'$ end. These sequences have been shown 
to be important for the propagation of narnaviruses, \cite{Est+03,Est+05}, but the mechanism 
which makes these termination sequences important has not been clear.

Finally, consider the evolution of a PolyProV2 system from a PolyProV1 virus. 
The reverse-complement recognition sequence would have to exist on the $5'$ end as a pre-requisite, but 
once this is in place, a partially ambigrammatic sequence can confer a partial advantage, 
so that the ambigrammatic property can evolve incrementally. 
There is no requirement for the reverse-read sequence to code for a functional protein. 
Beyond the requirement that there are no stop codons, there need not be any selective pressures on the 
reverse-read sequence. 

\section{Identification of PolyProV virus systems}
\label{sec: 3}

Next we consider the general principles which could be used to provide evidence 
for the existence of a PolyProV virus system. There are two approaches which could be used.

Because the defining feature of PolyProV viruses depends 
upon their interaction with ribosomes, ribosome profiling techniques should be important. 
In particular, we should address how these would distinguish ribosomes which have become 
\lq frozen' from those where translation is progressing. This 
approach could detect both PolyProV1 and PolyProV2 systems. 

The other approach is to use evidence from sequencing the viral RNA. 
The existence of ambigrammatic genes would be an indicator of a PolyProV2 system. 
(Because viruses undergo rapid mutations, the ambigrammatic property would not be conserved 
if it were not being used for some purpose, so that PolyProV1 systems are highly unlikely 
to be ambigrammatic). 
In this case we need to consider how to distinguish signatures of a PolyProV2 virus system 
from other possible explanations of ambigrammatic sequences.
 
 \subsection{Ribosome profiling}
 \label{sec: 3.1}
 
\begin{figure}
\begin{center}
\includegraphics[width=0.85\textwidth]{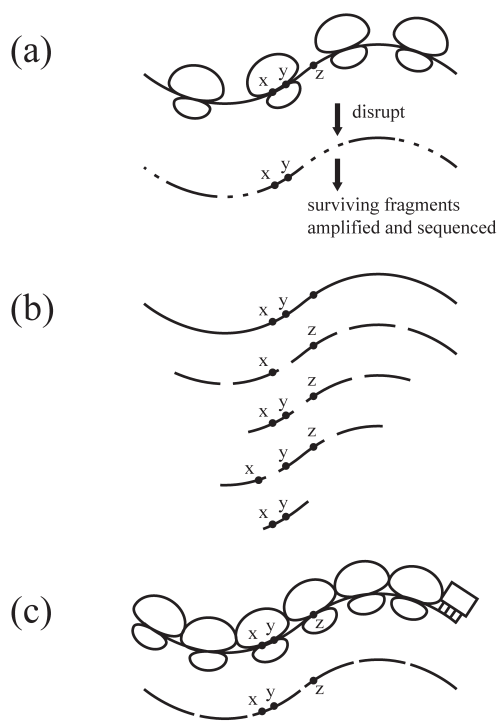}
\caption{\label{fig: 2}
({\bf a}) In ribosome profiling experiments, polysomes are disrupted and the RNA fragments which are under 
the \lq shadow' of a ribosome (approximately 30~nt long) are polymerised and sequenced. 
({\bf b}) If the ribosomes are moving along the polysome, the fragments containing a given base 
(indicated by $x$) will have that base positioned at any point within the fragment. Some of these fragments 
contain another base $y$, whereas others do not.
({\bf c}) If the polysomes are jammed, all of the fragments containing a given base will be very similar in structure, and 
have the base $x$ located at approximately the same position within the fragment. In this case the fragments that 
contain base $x$ almost always contain base $y$, but base $z$ is always found on a different fragment.
}
\end{center}
\end{figure}

\begin{figure}
\begin{center}
\includegraphics[width=0.85\textwidth]{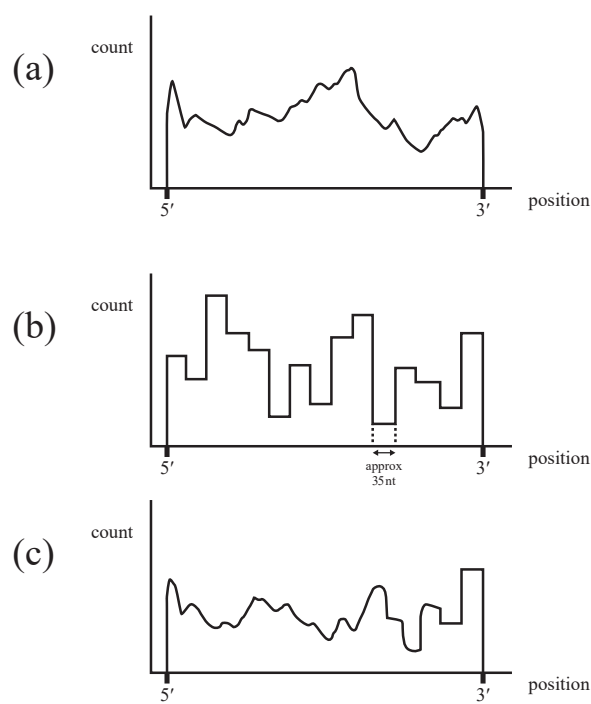}
\caption{\label{fig: 3}
({\bf a}) Illustrates the appearance of a typical ribosome profiling plot, observed when ribosomes 
are moving (left to right) along the RNA. ({\bf b}) If the ribosomes are frozen in fixed positions, the plot will 
show a sequence of plateaus. The width of each plateau is the length of the shadow of a ribosome, 
approximately 35~nt. ({\bf c}) If the separation of the jammed ribosomes fluctuates randomly as a function of time, 
the plateaus may become less distinct as we move away from the $3'$ end. 
}
\end{center}
\end{figure}

Ribosome profiling techniques~\cite{Ingolia2009,Ingolia2012} are based upon
mechanical disruption of polysome complexes formed by ribosomes and RNA,
followed by RNA sequencing. The mechanical disruption creates RNA fragments
which represent sections of the RNA strand which were covered by ribosomes
moment when the polysome was disrupted, as shown schematically in figure
\ref{fig: 2}({\bf a}) and ({\bf b}). The segments which lie under the \lq
shadow' of a ribosome are amplified and sequenced. These segments are
sufficiently long (approximately 30~nt) that their position in the genome can
be (in almost all cases) uniquely determined.  Ribosome profiling data is often
illustrated by plotting the frequency for counting fragments containing a base
$x$ as a function of the position of the base on the RNA chain.  Higher counts
are expected in regions where the ribosomes move more slowly, and a typical
ribosome profile plot has an appearance similar to the sketch in figure
\ref{fig: 3}({\bf a}). 

Consider how this technique can reveal jammed ribosomes, as indicated in figure \ref{fig: 1}(c). 
In order to understand the form of the expected profile, it is necessary to appreciate that the 
measured profile is a product of two factors: the desired signal, which is the ribosome coverage 
over a given nucleotide, must be multiplied by a factor which represents the amplification 
number of the RNA fragments. The latter is related to the fragment sequence in a manner
which is deterministic, but where the actual relationship is unknown. For this reason, the 
polymerisation amplification factor of a segment must be regarded as a random variable.

There is, however, one simple observation that we can make about the amplification number.
Because amplification involves successive replications of both the segment and its reverse 
complement, the amplification factor of the reverse complement of a segment must be 
highly correlated with that of the segment itself.
 
Now consider the ribosome profile resulting from stalled ribosomes. 
According to the PolyProV model, ribosomes will be prevented from detaching 
from the $3'$ end, and will form a close-packed array along the viral RNA, resembling 
a string of pearls.  Our model predicts that all of the ribosomes which are attached 
to the viral RNA would be located with their centres at quite narrowly defined 
positions on the RNA chain. If the ribosome profile were simply a reflection 
of the ribosome occupation at a locus, it would be constant. We should, however, 
take account of differences between the amplification factors of the segments.
Upon fragmentation, the region occupied by each ribosome 
would produce populations of similar fragments from all of the viral RNA molecules, as illustrated 
in figure \ref{fig: 2}({\bf c}). In particular, all of the bases under the shadow of a stalled ribosome 
are represented by the same population of RNA fragments, which have the same 
amplification factor. As we move along the chain, we encounter nucleotides 
which are under the shadow of an adjacent ribosome, and which are represented 
by a different RNA fragment, with a different amplification factor. 
At this point, the sequences which are being PCR amplified change abruptly. 
The replication rate of the new sequence is likely to be different, so that the 
heights of the plateaus will be different, forming an apparently random sequence, 
as illustrated in figure \ref{fig: 3}({\bf b}). The plateaus all have the same width, 
approximately 35~nt. This is in contrast to the results of ribosome profiling 
from an mRNA molecule which is being translated, where the there are many 
different fragments containing a given base. 

Our discussion of the ribosome profile of a PolyProV virus assumes that most of the 
ribosomes which are attached to viral RNA is stalled. The experimental data in \cite{Ret+20}
(figure 3) are very similar to the schematic illustration in figure \ref{fig: 3}{\bf b}. It is possible that 
some small fraction of the viral RNA is still being translated by moving ribosomes, while the \lq plateau' 
profile is visible, but the experiments suggest that most of the viral RNA which is in contact with 
ribosomes is in the \lq stalled' state. The experiments reported in \cite{Ret+20} suggest that only a small fraction 
of the viral RNA is bound to ribosomes, but that the bound fraction is mainly attached to stalled ribosomes. 
Current technology does not extend ribosome profiling to the single-cell level. If this becomes available, 
it may be possible to deduce how the viral infection progresses within a cell.

The separation of the frozen ribosomes may be subject to variations, because different base 
sequences bind to the ribosomes in a slightly different configuration. It is also possible that 
the length of the RNA strand which is covered by a ribosome might fluctuate as a 
function of time. These fluctuations would accumulate as we move further from the $3'$ end. 
In this case the plateaus in the ribosome profile plot would become less distinct as we move 
further from the $3'$ end, as illustrated in figure \ref{fig: 3}({\bf c}). 
The extent to which the ribosome profile plots would resemble 
figure \ref{fig: 3}({\bf c}) rather than figure \ref{fig: 3}({\bf b}) 
would have to be determined by experiment, but it would be expected to be a consistent feature 
of PolyProV systems.

\begin{figure}
\begin{center}
\includegraphics[width=0.85\textwidth]{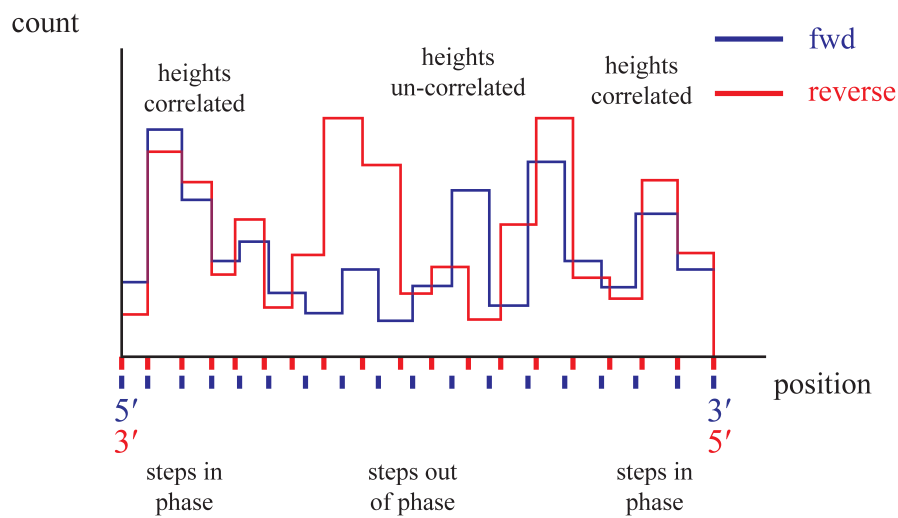}
\caption{\label{fig: 4}
In the case where both forward and complementary strands of a PolyProV2 system are detected 
in ribosome profiling, the plateaus on the complementary strands may overlap. If the lengths of 
the plateaus have some dependence upon the base sequence which is covered by the ribosome, 
the plateau widths vary apparently randomly, so that the forward and complementary strand plateaus 
are \lq in-phase' in some regions, and \lq out of phase' in others. The plateau heights are 
correlated when the strands are in phase.
}
\end{center}
\end{figure}

In the case of a PolyProV2 system, the ribosome profile plot for the reverse strands would also 
show a sequence of plateaus, which would be most distinct at the $5'$ end of the chain. 
If the plateaus in the profiles of both forward and complementary chains overlap, 
they could be \lq in phase', or \lq out of phase', or somewhere in between. 
For example, if the length of the region occupied by a ribosome is dependent on 
the sequence of bases that the ribosome covers, there will be apparently random 
variations in the lengths of the plateaus illustrated in figure \ref{fig: 3}({\bf b}), and 
the plateaus for the forward and complementary strands will be in phase over part of their length, 
and out of phase in other regions, as shown in figure \ref{fig: 4}. 
When the ribosome shadows of the forward and reverse strands 
are in phase, the segments which are amplified by the PCR process are complements of each other. 
When the plateaus are in phase, we expect the 
heights of the plateaus for the forward and complementary chains to be significantly 
correlated, because the polymerisation reaction involves multiple 
replications of \emph{both} forward \emph{and} complementary 
images of the fragments. In the regions where the ribosome shadows are 
out-of-phase, as in the centre section of the strand shown schematically in figure \ref{fig: 4}, the plateau heights of 
ribosome profiles from the forward and reverse strands will be uncorrelated.

 \subsection{Ambigrammatic sequences}
 \label{sec: 3.2}
 
 We have proposed that a PolyProV2 system could be detected by finding ambigrammatic viral genes 
 in sequencing studies. The detection of ambigrammatic sequences is an unambiguous 
 signal, and it is one which has already been observed in RNA virus sequences
 \cite{Cook2013,DeR+19,Cep20,Bat+20,Ret+20,Dudas2021}. It is necessary, however, 
 to consider whether alternative explanations are viable. 
 
 The possible explanations for observation of an ambigrammatic viral RNA sequence fall into 
 two classes. It might be that the reverse-readable sequences are expressed as proteins, which 
 serve some function in facilitating the propagation of the virus, for example, the protein 
 might poison defence mechanisms of the host cell, or it might form a complex with the viral 
 RNA which provides some protection. The other possibility is that the ambigrammatic property 
 provides some other advantage, without necessarily being expressed as a protein. The lack of stop codons 
 facilitates the association of ribosomes with the complementary RNA strand, so any plausible 
 mechanism would have to involve ribosomes in some way.  

There are three lines of evidence which can help to decide on the mechanism. The theory of the PolyProV2 
system is consistent with the evolution of the complementary strand sequence being neutral, because there 
is no role for the amino-acid sequence coded on the complementary chain (although some of the 
protein may be translated). One test of whether a sequence codes for a protein is to look at the 
ratio of non-synonymous to synonymous mutations which will be denoted by $R=\Delta N/\Delta S$, 
(where $\Delta N$ and $\Delta S$ are, respectively, the number of non-synonymous and synonymous mutations). 
We expect $R$ to be small when a readable base sequence is a functional gene coding for a protein, 
and the $R$ value for the forward sequence which codes for the RdRp is very small, indicating 
that this gene is strongly conserved. If both the forward and the complementary strands 
code for a protein, we might expect mutations which are synonymous for both forward and 
reverse transcription would be better tolerated. We shall report in detail upon
an investigation of this approach elsewhere~\cite{Dudas2021}. For both the 
\emph{Culex narnavirus 1} and \emph{Zheijiang mosquito virus 3}, the evidence indicates
that the complemenary strands of known ambigrammatic virus segments do not code for functional 
proteins.

Ambigrammatic sequences have been observed in a variety of simple RNA virus genomes
\cite{Cook2013,DeR+19,Cep20,Bat+20,Ret+20,Dudas2021}, 
but they are undoubtedly a rare phenomenon. Given that ambigrammatic sequences are rare, if 
two or more genes within a virus infection system are found to be ambigrammatic, this would be 
very unlikely to be the result of two functional genes arising on the complementary strand. 
An observation of the simultaneous detection of two or more ambigrammatic genes would 
strongly favour models, such as the PolyProV2 model, where there is an advantage in evolving an 
ambigrammatic sequence which is independent of whether the complementary strand open reading 
frames are translated into functional proteins.

Finally, finding some evidence for end recognition sequences would be an important 
part of validating the PolyProV model. It is reported that narnavirus sequences are 
typically terminated by CCCC at the 3' end and GGGG (an exact reverse complement) 
at the 5' end.  This observation suggests that CCCC and GGGG may be the recognition 
sequences, and that these are already present in many simple virus systems. 

\subsection{A candidate PolyProV2 system}
\label{sec: 3.3}

Recently, a mosquito-hosted narnavirus system (\emph{Culex narnavirus 1}) has been found to be associated with 
\emph{two} ambigrammatic genes \cite{Bat+20,Ret+20}. It has properties which make it a strong 
candidate to be a polysomal virus (see \cite{Bat+20,Ret+20} for a discussion of the experimental evidence):

\begin{enumerate}

\item There is a viral RNA segment which codes for the RdRp, and which resembles 
a narnavirus, but which has the property of being ambigrammatic, with forward and reverse codons 
aligned.  

\item Infection with this sequence is strongly associated with the presence of another RNA 
sequence, which was referred to in \cite{Bat+20} as the \lq Robin' sequence. 

\item The Robin sequence is also ambigrammatic over its entire length (about 850~nt), with the codons of the 
forward and reverse ORFs aligned. Neither forward nor reverse directions are homologous to known sequences.

\item Ribosome profiling experiments show a \lq plateau' structure
\cite{Ret+20}, which closely resembles that which is sketched in
figure~\ref{fig: 3}({\bf b}). The plateaus are seen in ribosome profiles of
both the RdRp gene and the Robin sequence. There is no evident loss of
definition of the plateaus on moving away from the $3'$ end, as illustrated in
figure~\ref{fig: 3}(c). This indicates that the packing of the ribosomes is
very tight.

\item The ribosome profile experiments detect the complementary strand of both the RdRp and the Robin sequence.
Both of the complementary strands have ribosome profiles with plateaus.

\item When the ribosome profiles of the forward and complementary strands are compared, the heights 
of the plateaus are correlated when they are in phase with each other, as illustrated 
in figure \ref{fig: 4}.

\item The companion and RdRp coding sequence share the feature of having complementary 
terminal sequences: both the RdRp and companion segments have one end terminating with CCCC, 
while the opposite end terminates GGGG.
 
\end{enumerate}

These features are consistent with the properties of a PolyProV2 type virus system, as described above.
In particular the fact that the two sequences are strongly correlated strongly implies that 
both are required for a viable infection. There is no evidence (in the form of 
overlapping fragments) that the two RNA molecules are 
ever found together as a single chain. There is also no evidence supporting the existence 
of any form of encapsulation of the two chains together. 
The observations are consistent with an infection by a system of two symbiotic viral RNA 
fragments. The natural hypothesis is that the Robin fragment is responsible for creating the 
molecule which blocks ribosome detachment. This fragment might encode a protein which has 
this role, or it might act directly in its RNA form with the RdRp-coding gene.

There is evidence that the reverse open reading frame is translated \cite{Ret+20}, although not into a 
functional protein \cite{Dudas2021}.
This is an overhead which does reduce the capacity of the cell to make functional viral proteins, and 
which would have to be balanced again whatever advantage arises from hiding both strands of the virus RNA.

Both the narnavirus component and the \lq Robin' segment contain GGGG and CCCC on their ends
suggesting, that the CCCC tetragram is the controlling switch to prevent detachment 
of the ribosome. The fact that these terminations are widely distributed in narnaviruses indicates that 
the ambigrammatic variants may be using a pre-existing feature as their 
recognition signal.

\section{Discussion}
\label{sec: 4}

We have proposed that viral RNA can be protected from degradation inside polysomes if these are \lq frozen'. 
This hypothesis explains recent observations \cite{Ret+20} of distinctive ribosome profiles of some narnaviruses. 
It also explains the existence of ambigrammatic sequences, because both phases of replication of 
an ambigrammatic gene can be protected. The use of protective polysome coverings may prove to be 
a widely distributed property of viral systems.

\section*{Author Contributions}
MW produced a draft of the manuscript following discussions with the other 
authors about the recent discovery of a narnavirus system which has two 
ambigrammatic genes. All authors contributed to writing the manuscript, and 
reviewed the manuscript before submission.

\section*{Acknowledgments}

We thank Hanna Retallack and Joe DeRisi for discussions of their
experimental studies of narnaviruses.  GH and DY were supported by the Chan
Zuckerberg Biohub; MW thanks the Chan Zuckerberg Biohub for its hospitality. 
We thank an anonymous referee for some interesting comments and suggestions, 
which were reflected in a revision of our manuscript.

\section*{References}

\bibliographystyle{iopart-num}

\bibliography{polyvirus}

\end{document}